\documentclass[11pt]{amsart}
\usepackage{amsmath,amsfonts,amssymb,amsthm}
\usepackage{amsrefs}
\usepackage{setspace}

\newcommand{\<}{\langle}
\renewcommand{\>}{\rangle}

\newcommand{\tr}{\operatorname{tr}}
\newcommand{\rank}{\operatorname{rank}}
\newcommand{\poly}{{\mathrm{poly}}} 

\newcommand{\G}{{\mathcal G}}
\renewcommand{\H}{{\mathcal H}}
\renewcommand{\S}{\mathcal{S}}

\newcommand{\R}{{\mathbb R}}
\newcommand{\Z}{{\mathbb Z}}
\renewcommand{\i}{\mathrm{i}}

\newtheorem{theorem}{Theorem}
\newtheorem{lemma}[theorem]{Lemma}

\mathchardef\ordinarycolon\mathcode`\:
\mathcode`\:=\string"8000
\def\vcentcolon{\mathrel{\mathop\ordinarycolon}} \begingroup
\catcode`\:=\active \lowercase{\endgroup \let :\vcentcolon }

\begin{document}

\title[Optimal measurements for the dihedral HSP]{Optimal measurements for the dihedral hidden subgroup problem}

\author{Dave Bacon}
\address{Santa Fe Institute, Santa Fe, NM 87501, USA}
\email{dabacon@santafe.edu}

\author{Andrew M. Childs}
\address{Institute for Quantum Information, 
California Institute of Technology,
Pasadena, CA 91125, USA}
\email{amchilds@caltech.edu}

\author{Wim van Dam}
\address{Department of Computer Science, University of California, Santa Barbara, Santa Barbara, CA 93106, USA}
\email{vandam@cs.ucsb.edu}


\begin{abstract}
We consider the dihedral hidden subgroup problem as the problem of
distinguishing hidden subgroup states.  We show that the optimal
measurement for solving this problem is the so-called pretty good
measurement.  We then prove that the success probability of this
measurement exhibits a sharp threshold as a function of the density
$\nu=k/\log_2 N$, where $k$ is the number of copies of the hidden
subgroup state and $2N$ is the order of the dihedral group.  In
particular, for $\nu<1$ the optimal measurement (and hence any
measurement) identifies the hidden subgroup with a probability that is
exponentially small in $\log N$, while for $\nu>1$ the optimal
measurement identifies the hidden subgroup with a probability of order
unity.  Thus the dihedral group provides an example of a group $\G$
for which $\Omega(\log|\G|)$ hidden subgroup states are necessary to
solve the hidden subgroup problem.  We also consider the optimal
measurement for determining a single bit of the answer, and show that
it exhibits the same threshold.  Finally, we consider implementing the
optimal measurement by a quantum circuit, and thereby establish
further connections between the dihedral hidden subgroup problem and
average case subset sum problems.  In particular, we show that an
efficient quantum algorithm for a restricted version of the optimal
measurement would imply an efficient quantum algorithm for the subset
sum problem, and conversely, that the ability to quantum sample from
subset sum solutions allows one to implement the optimal measurement.
\end{abstract}

\maketitle

\section{Introduction}

Quantum computers promise to solve certain problems asymptotically
faster than their classical counterparts.  In particular, Shor's
discovery of an efficient quantum algorithm for factoring
\cite{Shor:97}---a cryptographically significant task for which no
efficient classical algorithm is known---has motivated considerable
investigation into the potential algorithmic uses of quantum
computers.  Along with its predecessors
\cites{Deutsch:85a,Deutsch:92a,Bernstein:97a,Simon:97}, Shor's
algorithm can be viewed as a solution to one of a large class of
problems known as {\em hidden subgroup problems}
\cites{Boneh:95a,Hoyer:99a}, several of which also have applications to
interesting computational problems for which no efficient classical
algorithm is known.  Thus the broader question arises: under what
circumstances can the hidden subgroup problem (HSP) be solved
efficiently by a quantum computer?

Encouragingly, the quantum query complexity of the general HSP is
polynomial: for any group $\G$, only $\poly(\log|\G|)$ quantum queries
of the function that hides a subgroup are sufficient to solve the
problem \cites{Hoyer:99a,Ettinger:99a,Ettinger:99b,
Ettinger:00a,Ettinger:04a}.  Thus lower bounds showing that the HSP is
intractable are unlikely to be forthcoming.  However, the processing
of the queries could take exponential time, so it remains a challenge
to find algorithms that are efficient in terms of the number of
elementary operations.

Following Shor's discovery, there has been considerable progress in
showing that quantum computers can efficiently solve the HSP for
particular groups and particular kinds of subgroups
\cites{Boneh:95a,Kitaev:95a,Hales:99a,Hales:00a,Roetteler:98a,Grigni:04a,
Hallgren:00a,Ivanyos:03a,Friedl:03a,Moore:04a,Gavinsky:04a,IG:04}
(although there is also evidence that some hidden subgroup problems
may be hard even for quantum computers \cites{Kempe:04a,Moore:05a,Moore:05b}). 
However, for many hidden subgroup problems, the speedup offered by
quantum computers (if any) remains unknown.  In particular, no
efficient quantum algorithm is known for two cases whose applications
are of particular interest, the symmetric group and the dihedral
group.  For the former, an efficient quantum algorithm could be used
to efficiently solve the graph isomorphism problem
\cites{Boneh:95a,Beals:97a,Hoyer:97a,Ettinger:99b}, while for the
latter, an efficient quantum algorithm could be used to efficiently
solve certain cryptographically significant lattice problems
\cite{Regev:02a}.  Recent progress on the dihedral HSP has been
particularly encouraging: Kuperberg gave an algorithm using
subexponential (but superpolynomial) time and space
\cite{Kuperberg:03a}, and Regev improved this algorithm to use a
similar amount of time but only polynomial space \cite{Regev:04a}.

In this paper we concentrate on the dihedral hidden subgroup problem.
In particular, we study the optimal measurement for solving this
problem given samples of certain quantum states we call {\em hidden
subgroup states}.  We find that the success probability of the optimal
measurement exhibits a sharp threshold as a function of $k$, the
number of copies of the hidden subgroup state.  For the dihedral group
of order $2N$, let $k=\nu \log N$, where $\nu$ is the {\em density}.
(The logarithms in this article are always base $2$.)
For any fixed density $\nu>1$, the optimal measurement identifies the
hidden subgroup with constant probability, and therefore an efficient
quantum circuit for implementing this measurement would solve the
dihedral hidden subgroup problem.  (This can be compared to previous
results showing that a success probability of $1-{1/2N}$ can be
achieved with $\nu > 89$ \cite{Ettinger:00a}, and that a success
probability of $1/\poly(\log N)$ can be achieved with $\nu > 1$
\cite{Regev:02a}.) However, for any fixed $\nu<1$, the success
probability of the optimal measurement (and hence of any measurement)
is exponentially small in $\log N$.  This bound shows that
$\Omega(\log|\G|)$ hidden subgroup states are in fact necessary to
solve the dihedral HSP.  To the best of our knowledge this is the
first time more than a constant number of copies of the hidden
subgroup state have been shown to be necessary for any hidden subgroup
problem.

In addition to studying the success probability of the optimal
measurement, we also establish further connections between the
dihedral hidden subgroup problem and average-case subset sum problems
of density $\nu$.  Regev showed that the dihedral HSP can be solved
efficiently if one can efficiently solve average case subset sum
problems with $\nu>1$ \cite{Regev:02a}.  We show that the optimal
measurement for $k=\nu \log N$ copies can be implemented if one can
quantum sample from subset sum solutions at density $\nu$, and
conversely, that an implementation of the optimal measurement (of a
certain restricted form) by a quantum circuit can be used to solve the
average case subset sum problem.

Our results can be compared to those of Ip showing that Shor's
algorithm is an optimal solution to the abelian hidden subgroup
problem \cite{Ip:03a}.  In light of this observation, it is natural to
consider the optimal measurement for other hidden subgroup problems as
an approach to finding efficient algorithms.  Our results show that if
such an algorithm is efficient for the dihedral HSP, then one should
focus on finding an efficient quantum algorithm for the average case
subset sum problem with $\nu>1$ (or on implementing the measurement by
a quantum circuit not of the restricted form that could be used to
solve subset sum).

This paper is organized as follows.  In Section~\ref{sec:hsp}, we
review the hidden subgroup problem in general, and in
Section~\ref{sec:dhsp}, we review the dihedral hidden subgroup problem
in particular.  We present the optimal measurement for the dihedral
HSP in Section~\ref{sec:optimal} and establish bounds on its success
probability in Section~\ref{sec:sucprob}.  In Section~\ref{sec:info},
we show that the bounds of Section~\ref{sec:sucprob} are significantly
stronger than those one obtains from straightforward 
information-theoretic arguments.
Then, in Section~\ref{sec:onebit}, we show that the problem of
determining just the least significant bit of the answer requires
essentially as many copies of the hidden subgroup state as are
required to obtain the entire answer.  In Section~\ref{sec:subsetsum},
we establish connections between the optimal measurement for the
dihedral HSP and the subset sum problem.  Finally, we conclude in
Section~\ref{sec:discussion} with a discussion of the results and some
open problems.

\section{The hidden subgroup problem}
\label{sec:hsp}
We begin by reviewing the hidden subgroup problem.  Let $\G$ be a
finite group of order $|\G|$.  We assume that the elements of this
group can be efficiently represented as strings of $\poly(\log|\G|)$
bits.  Consider a function $f:\G \rightarrow S$ where $S$ is some
finite set whose elements can also be efficiently represented as
strings of $\poly(\log|\G|)$ bits.  In the HSP, we are given such a
function and promised that it is constant and distinct on left cosets
of some subgroup $\H \le \G$.  In other words,
$f(g_1)=f(g_2)$ if and only if $g_1$ and $g_2$ are in the same left coset of
$\H$.  The hidden subgroup problem is, given the ability to query the
function $f$, to produce a generating set for the subgroup $\H$.

In the quantum version of the HSP we are given a unitary operator
$U_f$ that computes the function $f$.  Explicitly, this quantum oracle
acts as
\begin{equation}
  U_f:|g,y\>  \mapsto  |g,y \oplus f(g)\>
\end{equation}
for all $g \in \G$ and $y \in S$, where $\oplus$ is the bitwise
exclusive or operation.  If we input the basis state $|g,0\>$ into
this oracle, it simply evaluates the function: $U_f |g,0\>=
|g,f(g)\>$.  Our goal is to use this black box to find generators of
the hidden subgroup in a time polynomial in $\log|\G|$.

In the standard approach to solving the hidden subgroup problem with a
quantum computer (used by all known quantum algorithms for the HSP),
one inputs a superposition over all group elements into the first
register and $|0\>$ into the second register, giving
\begin{equation}
   U_f :
   \frac{1}{\sqrt{|\G|}} \sum_{g \in \G} |g,0\>
    \mapsto  
   \frac{1}{\sqrt{|\G|}} \sum_{g \in \G} |g,f(g)\>
\,.
\end{equation}
Suppose we now discard the second register.  Due to the promise on
$f$, the state of the first register is then a mixed state whose form
depends on the hidden subgroup $\H$,
\begin{equation}
  \rho_\H  :=  \frac{|\H|}{|\G|} \sum_{g \in {\mathcal K}} |g\H\>\<g\H|
\label{eq:hss}
\end{equation}
where $\mathcal K \subset \G$ is a complete set of left coset
representatives of $\H$ in $\G$ (of size 
$|\mathcal K|=|\G|/|\H|$), and where we have defined the coset
states
\begin{equation}
  |g\H\>  :=  \frac{1}{\sqrt{|\H|}} \sum_{h \in \H} |gh\>
\,.
\end{equation}
We will call $\rho_\H$ the {\em hidden subgroup state} corresponding
to the subgroup $\H$.

Early quantum algorithms, including Deutsch's algorithm
\cite{Deutsch:85a}, the Deutsch-Jozsa algorithm \cite{Deutsch:92a},
the Bernstein-Vazirani algorithm \cite{Bernstein:97a}, Simon's
algorithm \cite{Simon:97}, and Shor's algorithm \cite{Shor:97}, all
solve examples of the abelian HSP but were not originally described in
this language.  The formulation in terms of a hidden subgroup was
presented by Boneh and Lipton \cite{Boneh:95a}, who also noted the
connection between the HSP over the symmetric group and the graph
isomorphism problem.

The HSP over arbitrary finite abelian groups has an efficient quantum
algorithm \cites{Simon:97,Shor:97,Boneh:95a,Kitaev:95a,Hales:00a}.
Hallgren, Russell, and Ta-Shma proved that the HSP has an efficient
quantum algorithm whenever the subgroup $\H$ is promised to be normal
and there is an efficient quantum Fourier transform over the group
$\G$ \cite{Hallgren:00a}.  Grigni, Schulman, Vazirani, and Vazirani
showed that the HSP over ``almost abelian'' groups has an
efficient quantum solution \cite{Grigni:04a}, and this result was
extended by Gavinsky to ``near-Hamiltonian'' groups
\cite{Gavinsky:04a}.  P{\"u}schel, R{\"o}tteler, and Beth gave an
efficient quantum algorithm for the HSP over the wreath product
$\Z_2^n \wr \Z_2$ \cite{Roetteler:98a}, and Friedl et al.\ showed how
to solve the HSP over a semidirect product $\Z_{p^k}^n \rtimes \Z_2$
for a fixed prime power $p^k$ \cite{Friedl:03a}.  Moore, Rockmore,
Russell, and Schulman gave an efficient quantum algorithm for the HSP
over certain semidirect product groups, the $q$-hedral groups
\cite{Moore:04a}, and Inui and Le Gall gave a solution for semidirect
product groups of the form $\Z_{p^k} \rtimes \Z_p$ with $p$ an odd
prime \cite{IG:04}.

Finally, as mentioned in the introduction, there is also a body of
knowledge about the query complexity of the HSP.  In particular,
Ettinger, H{\o}yer, and Knill have shown that $O(\log|\G|)$ quantum
queries of the function $f$ are sufficient to determine the hidden
subgroup \cite{Ettinger:99a}.  Unfortunately, the quantum algorithm
they present requires time $O(|\G|)$.

\section{The dihedral hidden subgroup problem}
\label{sec:dhsp}

The dihedral group of order $2N$, denoted ${\mathcal D}_N$, is the
group of symmetries of a regular $N$-sided polygon.  This group is
generated by two elements $r$ and $s$ satisfying the relations
$r^2=e$, $s^N=e$, and $rsr=s^{-1}$, where $e$ is the identity element.
Here $s$ corresponds to a rotation of the polygon and $r$ corresponds
to a reflection.  A generic element of the dihedral group can be
written as $r^t s^k$ where $t\in \Z_2$ and $s \in \Z_N$, and group
multiplication is given by $r^{t'} s^{k'} r^t s^k = r^{t+t'} s^{k
+(-1)^t k'}$.  (Throughout this article we write $\Z_N$ to denote
 $\Z/N\Z$.)

The dihedral hidden subgroup problem (DHSP) was first considered by
Ettinger and H{\o}yer \cite{Ettinger:00a}.  They showed that given
$O(\log N)$ queries to the hidden subgroup oracle for the dihedral
group, there exists a quantum algorithm whose output contains enough
classical information to solve the DHSP, and therefore that the query
complexity of the DHSP is $O(\log N)$.  Unfortunately, they were not
able to find an efficient algorithm to process the output, so this
approach has not yet led to an efficient algorithm for the DHSP.

A major motivation for attempting to solve the DHSP is a connection to
lattice problems discovered by Regev \cite{Regev:02a}.  A
$d$-dimensional lattice is the set of all integer linear combinations
of $d$ linearly independent vectors in $\R^d$ that form a basis for
the lattice.  In the {\em shortest vector problem}, one attempts to
find the shortest (nonzero) vector in the lattice given a basis.  In
particular, in the {\em $g(d)$ unique shortest vector problem}, we are
promised that the shortest vector is unique and shorter than all other
non-parallel vector by a factor $g(d)$.  The presumed hardness of
certain $g(d)$ unique shortest vector problems is the basis for a
cryptosystem proposed by Ajtai and Dwork (in which $g(d)=O(d^8)$)
\cite{Ajtai:97}, and a subsequent improvement proposed by Regev (in
which $g(d)=O(d^{1.5})$) \cite{Regev:03a}.  Regev showed that an
efficient quantum algorithm for the DHSP that works by sampling hidden
subgroup states can be used to solve the $\poly(d)$ unique shortest
vector problem \cite{Regev:02a}, thereby breaking the proposed lattice
cryptosystems.

Regev also gave a promising path toward solving the DHSP in the form
of a connection to the {\em subset sum problem}.  In the subset sum
problem, one is given $k$ numbers between $0$ and $N-1$, denoted $x
\in \Z_N^k$, and a target $t \in \Z_N$, and the goal is to find a
subset of the $k$ numbers, specified by a binary vector $b \in
\Z_2^k$, such that $b \cdot x = t$, where $b \cdot x := \sum_{j=1}^k
b_j x_j \bmod N$.  If such a subset exists, then we call $(x,t)$ a
legal subset sum input.  Regev has shown that if one can efficiently
solve $1/\poly(\log N)$ of the legal subset sum inputs (with $k>\log
N + 4$) then there is an efficient quantum algorithm for the DHSP
\cite{Regev:02a}.  While the general subset sum problem is NP-hard,
note that an algorithm for average-case inputs with $k > \log N + 4$
a fixed function of $N$ would be sufficient to solve the DHSP.

The first subexponential time quantum algorithm for the DHSP was given
by Kuperberg, who showed how to solve it in $2^{O(\sqrt{\log N})}$
time, space, and queries \cite{Kuperberg:03a}.  Regev reduced the
space requirement to $\poly(\log N)$ at the expense of only slightly
greater time and queries \cite{Regev:04a}.  Regev's approach also
shows a connection to the average case subset sum problem.

In trying to solve the DHSP, it is convenient to focus on a simplified
version that is in fact equivalent in difficulty to the full problem.
Specifically, we will focus on the case in which the subgroup $\H$ has
order two.  In general, there are two types of subgroups of ${\mathcal
D}_{N}$, cyclic subgroups and dihedral subgroups. The cyclic subgroups
consist only of rotations; they are of the form
\begin{equation}
  {\mathcal C}_{N/j}  :=  \{e,s^j,\dots,s^{-j}\}
\,,
\end{equation}
where $j \in \Z_N$ is a divisor of $N$.  Note that ${\mathcal C}_1$ is
simply the trivial subgroup. The dihedral subgroups consist of
rotations and reflections, and are of the form
\begin{equation}
  {\mathcal D}_{N/j,d}
     :=  \{e,s^j,\dots,s^{-j},rs^d,rs^{j+d}, \dots,rs^{-j+d}\}
\,,
\end{equation}
where $j \in \Z_N$ is a divisor of $N$ and $d \in \Z_N$.  Note that
${\mathcal D}_{N,d} = {\mathcal D}_N$ for any $d$.  Furthermore, note
that ${\mathcal D}_{1,d}=\{e,rs^d\}$ is an order two subgroup for any
$d$.  The cyclic subgroups ${\mathcal C}_{N/j}$ are all normal in
${\mathcal D}_N$ (that is, $ghg^{-1} \in {\mathcal C}_{N/j}$ for all
$h \in {\mathcal C}_{N/j}$ and $g \in {\mathcal D}_N$) while none of
the dihedral subgroups are normal except for the full dihedral group,
${\mathcal D}_N$.

Ettinger and H{\o}yer have shown that an efficient quantum algorithm
for the DHSP exists if one can solve the DHSP with the promise that
the hidden subgroup is either the trivial subgroup, ${\mathcal
C}_{1}=\{e\}$, or is some subgroup of order two, ${\mathcal
D}_{1,d}=\{e,rs^d\}$ for some (unknown) $d \in \Z_N$
\cite{Ettinger:00a}.  We will further restrict the problem by
determining the optimal measurement only for the order two subgroups.
In fact, it will turn out that when this restricted measurement
succeeds with high probability, it also identifies the trivial
subgroup with high probability, and therefore can be used to solve the
DHSP in general.

We will represent a dihedral group element $r^t s^k$ using two quantum
registers, $|t,k\>$, where the first register is a single qubit and
the second register consists of $\lceil \log N \rceil$ qubits.  When
the subgroup is an order two subgroup ${\mathcal D}_{1,d}$, then the
standard approach produces the random coset state
\begin{equation}
  |\phi_{k,d}\>  =  \frac{1}{\sqrt2} (|0,k\>+|1,-k+d\>)
\end{equation}
where $k$ is uniformly sampled from $\Z_N$ and addition is done in
$\Z_N$, i.e., modulo $N$.  In other words, the hidden subgroup state
corresponding to the subgroup $\H = {\mathcal D}_{1,d}$ is
\begin{equation}
  \rho_d  =  \frac{1}{N} \sum_{k \in \Z_N} |\phi_{k,d}\>\<\phi_{k,d}|
\,.
\end{equation}
It will be convenient to change the basis by Fourier transforming the
second register (over $\Z_N$) conditional on the first register being
$|1\>$ and inverse Fourier transforming the second register
conditional on the first register being $|0\>$.  In this new basis,
the hidden subgroup state is
\begin{equation}
  \rho_d  =  \frac{1}{N} \sum_{x \in \Z_N}
           |\tilde{\phi}_{x,d}\> \<\tilde{\phi}_{x,d}|
\end{equation}
where
\begin{equation}
  |\tilde{\phi}_{x,d}\>  :=  \frac{1}{\sqrt{2}}
                           (|0\>+\omega^{xd}|1\>) |x\>
\end{equation}
with $\omega:=\exp(2\pi \i/N)$.  When the subgroup is the trivial
group, the standard approach produces a random state $|t,x\>$ with $t$
uniformly sampled from $\Z_2$ and $x$ uniformly sampled from
$\Z_N$.  Thus the hidden subgroup state when the hidden subgroup is
$\H = {\mathcal C}_{1}$ is simply the maximally mixed state
\begin{equation}
  \rho_{\{e\}} = \frac{1}{N} \sum_{t \in \Z_2} \sum_{x \in \Z_N} |t,x\> \<t,x|
               = \frac{I_{2N}}{2N}
\end{equation}
where $I_{2N}$ is the $2N$-dimensional identity matrix.

Our goal is to determine $d$ given $k$ copies of the state $\rho_d$.
It will be helpful to write the state in a way that begins to reveal
the connection to the subset sum problem.  Note that we can write
\begin{equation}
  \rho_d  =  \frac{1}{2N} \sum_{b,c \in \Z_2} \sum_{x \in \Z_N}
           \omega^{(b-c)xd} |b,x\>\<c,x|
\,.
\end{equation}
Therefore
\begin{align}
  \rho_d^{\otimes k}
     &=  \frac{1}{(2N)^k} \sum_{b,c \in \Z_2^k} \sum_{x \in \Z_N^k}
       \omega^{[(b-c) \cdot x]d} |b,x\>\<c,x| \\
     &=  \frac{1}{(2N)^k} \sum_{x \in \Z_N^k} \sum_{p,q \in \Z_N}
       \omega^{d(p-q)} \sqrt{\eta_p^x \eta_q^x} |S_p^x,x\>\<S_q^x,x|
\label{eq:rho}
\end{align}
where $|S_r^x\>$ is the (normalized) uniform superposition over
subsets of $x \in \Z_N^k$ that sum to $r \in \Z_N$,
\begin{equation}
  |S_r^x\>  :=  \frac{1}{\sqrt{\eta_r^x}} \sum_{b \in S_r^x} |b\>
\end{equation}
with $S_r^x := \{ b \in \Z_2^k: b \cdot x = r \}$ denoting the set of
bit strings corresponding to subsets of $x$ that sum to $r$, and
$\eta_r^x := |S_r^x|$ denoting the number of such subsets.  If
$\eta^x_r=0$, then no such state can be defined, and we use the
convention $|S^x_r\>=0$.  When the subgroup is trivial, $k$ copies of
the hidden subgroup state are simply $k$ copies of the maximally mixed
state, $\rho_{\{e\}}^{\otimes k}=I_{(2N)^k}/(2N)^k$.

\section{The optimal measurement}
\label{sec:optimal}

In this section, we present the optimal measurement for distinguishing
the hidden subgroup states $\rho_d^{\otimes k}$.  The measurement will
have $N$ outcomes, one for each possible value of $d$, and will be
optimal in the sense that the probability of obtaining the correct
outcome will be as large as possible.  Recall that a general quantum
measurement, a positive operator-valued measure (POVM), is specified
by a set of positive operators $\{ E_j \}$, $E_j > 0$, that sum to the
identity, i.e., $\sum_j E_j=I$.  Given a density matrix $\rho$, the
probability of obtaining the outcome $j$ is $\tr E_j \rho$.

Ip was the first to consider optimal measurements for
hidden subgroup problems \cite{Ip:03a}.  In particular, he found the optimal
measurement for the abelian hidden subgroup problem when the hidden
subgroups are all given with equal a priori probabilities, thereby
showing that the methods developed to solve the factoring problem are
optimal.  Ip also derived the optimal measurement for the dihedral
hidden subgroup problem given a single copy of the hidden subgroup
state.  As we shall see, this measurement fails to efficiently
identify the order two subgroups.  Since we know that there exists a
measurement for solving the DHSP using $O(\log N)$ copies of the
hidden subgroup states, it is of interest to understand the optimal
measurement given $k \gg 1$ copies.

The optimal measurement turns out to be the {\em pretty good
measurement} (PGM) \cite{HW:94} (also known as the {\em square root
measurement} or {\em least squares measurement}) \cite{Eldar:02}.

To prove that the PGM is optimal, we will use the following theorem:
\begin{theorem}[Holevo \cite{Holevo:73b}, Yuen-Kennedy-Lax
\cite{Yuen:75}]
\label{thm:holevo}
Given an ensemble of quantum states $\rho_i$ with a priori
probabilities $p_i$, the measurement with POVM elements $E_j$ maximizes
the probability of successfully identifying the state if and only if
\begin{align}
  \sum_i p_i \rho_i E_i &= \sum_i p_i E_i \rho_i \label{eq:holevons1} \\
  \text{and}\quad
  \sum_i p_i \rho_i E_i &\ge p_j \rho_j  \quad \forall j
  \label{eq:holevons2}
\,.
\end{align}
\end{theorem}
\noindent
This condition follows most easily from noting that the maximization
problem is a semidefinite program \cites{Yuen:75,Eldar:03a,Ip:03a}.
While Theorem~\ref{thm:holevo} provides necessary and sufficient
conditions for a measurement to be optimal, it is nontrivial in
general to construct measurements that satisfy these conditions.

Given $\rho_d^{\otimes k}$ with equal a priori probabilities for each
$d \in \Z_N$, we wish to find the measurements $\{E_j\}_{j \in \Z_N}$
that maximize the
probabilities of correctly identifying these states, where we identify
the measurement outcome $j$ with our guess for the hidden subgroup
label $d$.  The PGM is given by
\begin{equation}
  E_j  =  G^{-1/2} \rho_j^{\otimes k} G^{-1/2}
\label{eq:pgm}
\end{equation}
where the inverse is taken over the support of $G$, and where
\begin{align}
  G  &:=  \sum_{j \in \Z_N} \rho_j^{\otimes k} \\
     & =  \frac{1}{(2N)^k} \sum_{x \in \Z_N^k} \sum_{p,q \in \Z_N}
        \sum_{j \in \Z_N} \omega^{j(p-q)} \sqrt{\eta_p^x \eta_q^x}
        |S_p^x,x\>\<S_q^x,x| \\
     & =  \frac{N}{(2N)^k} \sum_{x \in \Z_N^k} \sum_{r \in \Z_N}
        \eta_r^x |S_r^x,x\>\<S_r^x,x|
\,.
\label{eq:g}
\end{align}
Inserting (\ref{eq:rho}) and (\ref{eq:g}) into (\ref{eq:pgm}), we find
that the pretty good measurement for the dihedral hidden subgroup
states has the measurement operators
\begin{equation}
    E_j  =  \frac{1}{N} \sum_{x \in \Z_N^k} \sum_{p,q \in \Z_N}
         \omega^{j(p-q)} |S_p^x,x\>\<S_q^x,x|
\,.
\label{eq:povm}
\end{equation}

That this measurement is optimal can be seen by substitution into
(\ref{eq:holevons1}) and (\ref{eq:holevons2}).  We have
\begin{equation}
  \sum_{i \in \Z_N} \rho_i E_i
     =  \frac{1}{(2N)^k} \sum_{x \in \Z_N^k} \sum_{p,q \in \Z_N}
      \sqrt{\eta_p^x \eta_q^x} |S_p^x,x\>\<S_p^x,x|
     = \sum_{i \in \Z_N} E_i \rho_i
\end{equation}
which verifies (\ref{eq:holevons1}).  Now $\rho_j$ is block diagonal
with rank one blocks: $\rho_j = \sum_{x \in \Z_N^k}
|\rho^x_j,x\>\<\rho^x_j,x|$ where
\begin{equation}
  |\rho^x_j\> = \frac{1}{(2N)^{k/2}} \sum_p \omega^{jp}
                \sqrt{\eta^x_p} |S^x_p\>
\,.
\end{equation}
For any $j \in \Z_N$ and for any block $x \in \Z_N^k$, we find
\begin{align}
  \<\rho^x_j,x| \sum_{i \in \Z_N} \rho_i E_i |\rho^x_j,x\>
     &= \frac{1}{(2N)^k} \sum_{p,q \in \Z_N}
        \eta_p^x \sqrt{\eta_p^x \eta_q^x} \\
     &\ge \frac{1}{(2N)^k} \sum_{p \in \Z_N} \eta^x_p
      = \<\rho^x_j|\rho^x_j\>
\,,
\end{align}
which verifies (\ref{eq:holevons2}).

Notice that since $G$ is not supported on the entire
$(2N)^k$-dimensional space, the operators $\{E_j\}_{j \in \Z_N}$ do
not form a complete partition of the identity.  (The dimension of the
support of $G$, $\rank G = |\{(x,p): x \in \Z_N^k, p \in \Z_N,
\eta^x_p > 0\}|$, is given by Sloane's integer sequence A098966
\cite{Sloane:04}.)  To complete the measurement we can add an
additional measurement operator, $E_{\{e\}}:=I-\sum_{j\in \Z_N} E_j$.
We associate this measurement outcome with the trivial subgroup.  We
emphasize that the measurement is only optimized for determining the
order two subgroups.  However, we will see that the optimal measurement
with the additional measurement operator $E_{\{e\}}$ also efficiently
identifies the trivial subgroup.  Of course, the optimal measurement
for the full dihedral group is never any better than the optimal
measurement for distinguishing the order two subgroups.

\section{Success probability of the optimal measurement}
\label{sec:sucprob}

In this section we study the success probability of the optimal
measurement for distinguishing the dihedral hidden subgroup states.
Using the expressions (\ref{eq:rho}) and (\ref{eq:povm}), a simple
calculation shows that the probability of successfully identifying an
order two subgroup is independent of the hidden shift $d$ and is given
by
\begin{equation}
  p  :=  \tr E_d \rho_d^{\otimes k}  \\
     =   \frac{1}{2^k N^{k+1}}
        \sum_{x \in \Z_N^k}
        \bigg( \sum_{r \in \Z_N} \sqrt{\eta_r^x} \bigg)^2
\,.
\label{eq:sucprob}
\end{equation}

We now show that the success probability has a sharp threshold as a
function of the density $\nu=k/\log N$.  More precisely, we find

\begin{theorem}
\label{thm:threshold}
If $\nu \ge 1 + \frac{4}{\log N}$, then the probability of
successfully determining the order two subgroup is at least $1/8$.
Furthermore, for any $N$ and $k$, the probability of successfully
determining the order two subgroup is less than $2^k/N$ (which in
particular is exponentially small in $\log N$ for any fixed $\nu <
1$).
\end{theorem}

We will need the following lemma to prove the first statement of the
theorem.
\begin{lemma}
[Cf.\ proof of Lemma~4.1 of \cite{Regev:02a}]
\label{lem:regev}
For fixed $r \in \Z_N$ and uniformly random $x\in \Z_N^k$,
\begin{equation}
  \Pr\bigg(\eta^x_r  \ge  \frac{2^k-1}{2N}\bigg) \ge 1 - \frac{4N}{2^k - 1}
\,.
\end{equation}
\end{lemma}
\noindent
With this fact in hand, we can establish our main result.

\begin{proof}[Proof of Theorem~\ref{thm:threshold}]
For the lower bound on the success probability, we have
\begin{equation}
  p  \ge  \frac{1}{2^k N} \bigg(
        \frac{1}{N^k} \sum_{x \in \Z_N^k} \sum_{r \in \Z_N}
        \sqrt{\eta_r^x} \bigg)^2
\end{equation}
by Cauchy's inequality applied to (\ref{eq:sucprob}).  Now by
Lemma~\ref{lem:regev},
\begin{align}
  \frac{1}{N^k} \sum_{x \in \Z_N^k} \sqrt{\eta^x_r}
   & \ge  \sqrt\frac{2^k-1}{2N} \,
       \Pr\bigg(\eta^x_r \ge \frac{2^k-1}{2N}\bigg) \\
   & \ge  \sqrt\frac{2^k-1}{2N} - \sqrt\frac{8N}{2^k-1}
\end{align}
for any $r$, which implies
\begin{align}
  p  &\ge  \frac{N}{2^k} \bigg( \sqrt\frac{2^k-1}{2N} -
                              \sqrt\frac{8N}{2^k-1} \bigg)^2 \\
     &\ge  \frac{2^k-1}{2^{k+1}} - \frac{4N}{2^k} \\
     &\ge  \frac{1}{4} - \frac{1}{2^k} \\
     &\ge  \frac{1}{8}
\end{align}
where we have assumed $k \ge \log N + 4$ (and also, in particular,
we have used $k \ge 3$).

For the upper bound on the success probability, we have
\begin{align}
  p  &\le  \frac{1}{2^k N^{k+1}} \sum_{x \in \Z_N^k}
         \bigg( \sum_{r \in \Z_N} \eta_r^x \bigg)^2 \\
     &=    \frac{2^k}{N}
\end{align}
where in the first line we have used the fact that the $\eta$'s are
all integers to remove the square root in (\ref{eq:sucprob}), and in
the second line we have used the fact that $\sum_{r \in \Z_N} \eta_r^x
= 2^k$ for any $x$.  This completes the proof.
\end{proof}

We claimed earlier that when the measurement identifies the order two
subgroups with reasonable probability, it will also identify the
trivial subgroup.  This follows from a simple calculation: supposing
$\nu \ge 1 + \frac{4}{\log N}$,
\begin{align}
  p_{\{e\}}  &:=   \tr E_{\{e\}} \rho_{\{e\}}^{\otimes k} \\
             &=    1 - \frac{\rank G}{(2N)^k} \\
             &\ge  1 - \frac{N}{2^k} \\
             &\ge  \frac{15}{16}
\,.
\end{align}

\section{Bounds by information-theoretic arguments}
\label{sec:info}

The proof of the above threshold theorem used specific properties of
the dihedral hidden subgroup problem.  It is reasonable to ask if this
is necessary, or if one could instead obtain the same bounds using the
 powerful techniques of quantum information theory.  This appears not
to be the case.  Here we derive the 
 information-theoretic lower bound bound
$\nu = k/\log N \ge p$, which is weaker than the $\nu \ge 1$ bound
of Theorem~\ref{thm:threshold} for probabilistic, non-exact
algorithms. 

Given $k$ copies of the hidden subgroup state $\rho_d$, we want to
determine the outcome $d$ with success probability at least $p$.
Viewed in a data transmission setting, we can imagine a sender
encoding $\log N$ bits of information (the value $d\in\Z_N$) in the
quantum state $\rho_d^{\otimes k}$, after which a receiver decodes the
$\log N$ bits by solving the DHSP using the $k$ copies of $\rho_d$.
The number of copies $k$ required for this approach to work can be
analyzed with the tools of quantum information theory, thereby giving
a lower bound on $k$.  Because the amount of information received
depends on the success probability $p$, the lower bound on $k$ will
also depend on $p$.  Roughly speaking, the amount of information that
can be transmitted with $k$ copies is upper bounded by $k$ bits, while
the received amount of information is lower bounded by $p \log N$,
leading to the lower bound $k \ge p \log N$.  The details are as
follows.

Given $N$, we define a {\em source} $\S_k$ that draws from the uniform
ensemble $\{(1/N,\rho_d^{\otimes k})\}_{d \in \Z_N}$ where each $d$
occurs with equal probability $1/N$.  Holevo's $\chi$ quantity,
defined by
\begin{equation}
\chi(\S_k)  :=  S\Big(\frac{1}{N} \sum_d \rho_d^{\otimes k}\Big) -
\frac{1}{N}\sum_{d} S(\rho_d^{\otimes k})
\end{equation}
where $S(\cdot)$ denotes the Von Neumann entropy of a mixed quantum
state \cite{NielsenChuang}, gives an upper bound on the accessible
information of the ensemble.  The state $\rho_d$ is defined in a
$2N$-dimensional Hilbert space, and its spectrum consists of the
eigenvalues $1/N$ and $0$, each with multiplicity $N$, while the
spectrum of the mixture $\frac{1}{N} \sum_d \rho_d$ consists of the
eigenvalues $1/N$ and $0$, each with multiplicity $1$, and the
eigenvalue $1/2N$ with multiplicity $2N-2$.  Hence, for the $k=1$ case,
we have $\chi(\S_1) = 1-1/N$.  For general $k$, we have
$S(\rho_d^{\otimes k}) = k \log N$ since the entropy is additive under
tensor products.  For the mixture $\frac{1}{N}\sum \rho_d^{\otimes
k}$ we note that the reduced density matrices of each copy 
are equal to $\frac{1}{N}\sum \rho_d$. 
Hence, by subadditivity of the Von Neumann entropy, the 
entropy of  this mixture
 is bounded from above by $k\,S(\frac{1}{N}\sum_d{\rho_d})$.
  Overall, this implies an upper bound of $k(1-1/N)$ on the
accessible information of $\S_k$.

Now, on the receiver's end, if the message can be decoded without
error, then $\S_k$ has a capacity of $\log N$ bits per message.
However, we should take into account that we are satisfied with a
constant success probability $p$, which can be smaller than $1$.  In
this more general case, the information transmitted by the source will
be bounded from below by $I_p \ge \log N -
H(p,\frac{1-p}{N-1},\ldots,\frac{1-p}{N-1})$, where $H(\cdot)$ is the Shannon
entropy of a probability distribution.  Since $\chi(\S_k) \ge I_p$, we
find
\begin{align}
k\left({1-\frac{1}{N}}\right)  
& \ge  \log N - H\left({p,\frac{1-p}{N-1},\dots,\frac{1-p}{N-1}}\right)  \\
&  =   \log N -(1-p)\log(N-1) - H(p,1-p)\,.
\end{align}
For constant $p$ and large $N$, this converges to the lower bound $k
\geq p \log(N-1) - H(p,1-p)$, which is significantly weaker than the
bound $k \geq \log N$ from our earlier Theorem~\ref{thm:threshold}.

\section{Determining the least significant bit}
\label{sec:onebit}

Although the determination of the entire shift $d$ requires at least
$\log N$ copies of the hidden subgroup state, one might hope to
acquire partial information about the shift using fewer copies.  For
example, suppose one could determine the least significant bit of the
shift using only a single hidden subgroup state.  An iterative
determination of the entire shift using such a measurement as a
subroutine would still require $\log N$ hidden subgroup states, but
the basic measurement for determining a single bit would be much
simpler.  However, here we rule out such a possibility: the optimal
measurement for determining even just a single bit of the shift still
requires $\log N$ hidden subgroup states.  More precisely, we prove
the following:

\begin{theorem}
\label{thm:onebit}
  With $k = \nu \log N$ copies of the dihedral hidden subgroup state
  $\rho_d$, the probability of successfully identifying the least
  significant bit of $d$ is exponentially close to $\frac{1}{2}$ for
  any fixed $\nu < 1$.
\end{theorem}
\noindent
Note that since there is a measurement to determine the entire shift
with constant probability for any fixed $\nu > 1$, in particular there
is a measurement to determine the least significant bit with
probability bounded away from $1/2$ in this regime.  Thus
Theorem~\ref{thm:onebit} shows that the threshold for success remains
essentially the same, at $\nu \sim 1$, for the problem of determining
just the least significant bit.

To establish this result, we proceed as before: we first identify the
optimal measurement, then derive an expression for its success
probability, and finally place bounds on this expression.  Our goal is
to determine the least significant bit of $d$, i.e., whether $d$ is
even or odd.  In other words, we would like to distinguish the two
density matrices
\begin{equation}
  \rho_\pm  := 
    \frac{2}{N} \sum_{d \text{~even,odd}} \rho_d^{\otimes k}
\,.
\end{equation}
Since $\rho_+ + \rho_- = \frac{2}{N} G$ (with $G$ given in
(\ref{eq:g})), the PGM for these states has the two POVM operators
\begin{align}
  E_\pm  &:=  \frac{N}{2} G^{-1/2} \rho_\pm G^{-1/2}\\
        &= \sum_{d \text{~even,odd}} E_d
\,.
\end{align}
Now for simplicity, we assume $N$ is even.  The identity
\begin{equation}
  \sum_{d \text{~even}} \omega^{d(p-q)}
   =  \begin{cases}
      \frac{N}{2}&   p=\pm q \\
      0          &   \text{otherwise}
    \end{cases}
\end{equation}
can then be used to simplify these expressions, and we obtain
\begin{align}
  \rho_\pm  &=  \frac{1}{(2N)^k} \sum_{x \in \Z_N^k} \sum_{r \in \Z_N}
              (\eta^x_r |S^x_r,x\>\<S^x_r,x| \pm
              \sqrt{\eta^x_r \eta^x_{-r}} |S^x_r,x\>\<S^x_{-r},x|) \\
     E_\pm  &=  \frac{1}{2} \sum_{x \in \Z_N^k} \sum_{r \in \Z_N}
              (|S^x_r,x\>\<S^x_r,x| \pm |S^x_r,x\>\<S^x_{-r},x|)
\,.
\end{align}

We claim that this PGM is the optimal measurement for determining the
least significant bit of the shift.  To see this, check the conditions
of Theorem~\ref{thm:holevo}.  We have
\begin{align}
  \sum_{i \in \Z_N} \rho_i E_i
     &=  \rho_+ E_+ + \rho_- E_- \\
     &=  \frac{1}{(2N)^k} \sum_{x \in \Z_N^k} \sum_{r \in \Z_N}
       \Big(\eta^x_r + \sqrt{\eta^x_r \eta^x_{-r}}\Big)
       |S^x_r,x\>\<S^x_r,x|
\end{align}
since the cross terms cancel.  A similar calculation verifies
(\ref{eq:holevons1}).  Then

\begin{equation}
   \sum_{i \in \Z_N} \!\! \rho_i E_i \!-\! \rho_\pm
   \!=\!   \frac{1}{(2N)^k} \!\! \sum_{x \in \Z_N^k} \! \sum_{r \in \Z_N} \!\!\!
     \sqrt{\eta^x_r \eta^x_{-r}} (|S^x_r,x\>\<S^x_r,x| \!\mp\!
                                  |S^x_r,x\>\<S^x_{-r},x|)
,
\end{equation}
which is clearly a positive matrix, verifying (\ref{eq:holevons2}).

The success probability of this optimal measurement is independent of
whether $d$ is even or odd, and is given by
\begin{align}
   \tilde p  &:=  \tr E_+ \rho_+ \\
             &=   \frac{1}{2(2N)^k} \!\! \sum_{x \in \Z_N^k} \!
                  \sum_{r \in \Z_N} \!\!
                \Big(\eta^x_r \!+\! \sqrt{\eta^x_r \eta^x_{-r}}\Big)
                \tr(|S^x_r,x\>\<S^x_r,x| \!+\! |S^x_r,x\>\<S^x_{-r},x|) \\
             &=   \frac{1}{2}\Big[1 + \frac{1}{(2N)^k} \Big(
                \sum_{x \in \Z_N^k} \sum_{r \in \Z_N}
                \sqrt{\eta^x_r \eta^x_{-r}} + 2 \eta^x_0 + 2
                \eta^x_{N/2} \Big)\Big]
\label{eq:onebitsucprob}
\end{align}
where we have used the fact that $\sum_{x \in \Z_N^k} \sum_{r \in
\Z_N} \eta^x_r = (2N)^k$.

With these expressions in hand, we are ready to prove
Theorem~\ref{thm:onebit}:

\begin{proof}
We bound the expression (\ref{eq:onebitsucprob}) for the success
probability of the optimal measurement.  First consider the cases
$r=0,N/2$.  For $r=0$ we have
\begin{align}
  \frac{1}{(2N)^k} \sum_{x \in \Z_N^k} \eta^x_0
     &=  \frac{N^{k-1} (2^k-1) + N^k}{(2N)^k} \\
     &\le  \frac{1}{N} + \frac{1}{2^k}
\end{align}
and for $r=N/2$ we have
\begin{align}
  \frac{1}{(2N)^k} \sum_{x \in \Z_N^k} \eta^x_{N/2}
     &=   \frac{N^{k-1} (2^k-1)}{(2N)^k} \\
     &\le  \frac{1}{N}
\,.
\end{align}
In fact, the latter expression holds for any $r \ne 0$, since for any
non-empty subset of numbers, specifying all but one of those numbers
leaves exactly one possible subset summing to $r$.  The additional
$N^k$ term for $r=0$ comes from the contribution of the empty set for
each of the $N^k$ possible assignments of the $x$'s.

For the remaining terms, we have
\begin{align}
  \frac{1}{(2N)^k} \sum_{r \ne 0,N/2} \sum_{x \in \Z_N}
    \sqrt{\eta^x_r \eta^x_{-r}}
     &\le 
  \frac{1}{(2N)^k} \sum_{r \ne 0,N/2} \sum_{x \in \Z_N}
    \eta^x_r \eta^x_{-r} \\
     &= 
  \frac{(N-2)(2^k-1)(2^k-2)N^{k-2}}{(2N)^k} \\
     &\le 
  \frac{2^k}{N}
\,.
\end{align}
In the first line we have used the fact that the $\eta$'s are integers
to remove the square root. In the second line we consider fixing one
of the $N-2$ values of $r$, and consider a non-empty subset $S$ (of
which there are $2^k-1$) and a distinct non-empty subset $T$ (of which
there are $2^k-2$).  If we consider two elements $i,j$ such that
either $i \in S-T$ and $j \in T$, or $i \in T-S$ and $j \in S$ (such a
choice is always possible because $S$ and $T$ are non-empty and
distinct), then for any values of the remaining $k-2$ elements, there
is exactly one choice for elements $i$ and $j$ such that the elements
in $S$ sum to $r$ and the elements in $T$ sum to $-r$.  Thus the sum
is exactly $(N-2)(2^k-1)(2^k-2)N^{k-2}$.

Using these expressions in (\ref{eq:onebitsucprob}), we find
\begin{equation}
  \tilde p  \le  \frac{1}{2}
    \Big( 1 + \frac{2^k}{N} + \frac{6}{N} + \frac{3}{2^k} \Big)
\,.
\end{equation}
Thus we see that with $k = \nu \log N$, the success probability is
exponentially close to $1/2$ for any fixed $\nu < 1$.
\end{proof}

\section{Relation to the subset sum problem}
\label{sec:subsetsum}

Given that the optimal measurement solves the DHSP if (and only if)
$\nu > 1$, we would like to understand whether this measurement can be
implemented efficiently.  In this section we consider how to implement
the measurement by a quantum circuit, and we find that its
implementation is closely related to the subset sum problem.

Recall the definition of the subset sum problem: given $x \in \Z_N^k$
and $t \in \Z_N$, find a subset $b \in \Z_2^k$ such that $b \cdot x =
t$.  If such a $b$ exists, we call $(x,t)$ a legal instance.  In the
decision version of the subset sum problem, we wish to determine only
whether a given instance is legal or not.  This problem is
NP-complete.  We might also want to return one or more of the subsets
$b$ in the case where the instance is legal.  Regev has shown that if
there exists an efficient algorithm for finding one such subset for a
large fraction of the legal instances, then one could solve the
dihedral hidden subgroup problem efficiently \cite{Regev:02a}.  More
precisely,

\begin{theorem}[Regev \cite{Regev:02a}]
\label{thm:regev}
If there exists an efficient algorithm that finds a subset $b$ such
that $b \cdot x = t$ for a fraction $1/\poly(\log N)$ 
of the legal
subset sum instances $(x,t)$ when $k>\log N+4$, then there exists an
efficient quantum algorithm for the dihedral hidden subgroup problem.
\end{theorem}

Here we show a similar result for the implementation of the optimal
measurement for the dihedral hidden subgroup states.  Namely, if one
can efficiently quantum sample from subset sum solutions at density
$\nu = k/\log N$, then one can efficiently implement the optimal
measurement for the DHSP with $k$ copies.  We also show a weak
converse to this result: if one can efficiently implement the optimal
measurement by a quantum circuit (under a certain restriction), then
one can in turn solve the average case subset sum problem of
corresponding density (and indeed, can quantum sample from subset sum
solutions).

Recall from (\ref{eq:povm}) that the POVM operators for the 
DHSP can be expressed as
\begin{align}
    E_j  &=  \frac{1}{N} \sum_{x \in \Z_N^k} \sum_{p,q \in \Z_N}
         \omega^{j(p-q)} |S_p^x,x\>\<S_q^x,x| \\
   &=  \sum_{x \in \Z_N^k} E_j^x \otimes |x\>\<x|
\end{align}
where
\begin{equation}
  E_j^x  :=  \frac{1}{N} \sum_{p,q \in \Z_N} \omega^{j(p-q)}
           |S_p^x\>\<S_q^x|
\,.
\label{eq:povmwithx}
\end{equation}
In other words, each $E_j$ is block diagonal, with blocks labeled by
some $x \in \Z_N^k$.  Because each $E_j$ has high rank, there is
considerable freedom in how one implements the measurement by a
quantum circuit.  However, from a representation-theoretic perspective
(see the Appendix), it is natural to perform this measurement in a
particular way, first measuring the label $x$ and then performing the
POVM $\{E_j^x\}_{j \in \Z_N}$ conditioned on that label.  Note that
each $E_j^x$ is rank one, so that the POVM $\{E_j^x\}_{j \in \Z_N}$
for fixed $x \in \Z_N^k$ is refined into one-dimensional subspaces,
removing much of the freedom in the implementation of the original
POVM $\{E_j\}$.

For any given $x \in \Z_N^k$, we consider the implementation of the
POVM $\{E_j^x\}_{j \in \Z_N}$ by an $x$-dependent quantum circuit
followed by a measurement in the computational basis to give the
outcome $j$.  In general, this circuit and measurement will act on a
larger Hilbert space than is required to hold the original input.  The
quantum circuit will then correspond to some unitary operation $U^x$
on the larger space.  Without loss of generality, we can assume that
the final measurement is in a basis $\{|j\>\}$ such that the values $j
\in \{0,1,\ldots,N-1\}$ indicate the measurement outcome $E_j$.
According to Neumark's theorem, the unitary operator $U$ has the block
form
\begin{equation}
  U^x  =  \begin{pmatrix}
          V^x &  A^x \\ 
          B^x &  C^x 
        \end{pmatrix}
\end{equation}
where
\begin{equation}
  V^x  :=  \frac{1}{\sqrt N} \sum_{j,q \in \Z_N} \omega^{-jq} |j\>\<S^x_q|
\end{equation}
is a fixed $(N \times 2^k)$-dimensional matrix whose columns are the
(subnormalized) vectors corresponding to the rank one POVM elements
$\{E^x_j\}$, and $A^x,B^x,C^x$ are arbitrary up to the requirement
that $U^x$ is unitary.  It is convenient to perform a Fourier
transform on the left, i.e., on the index $j$ (over $\Z_N$, for the
relevant values $j \in \{0,1,\ldots,N-1\}$), giving a unitary operator
\begin{equation}
  \tilde U^x  =  \begin{pmatrix}
                 \tilde V^x & A^x \\ 
                 \tilde B^x & C^x 
               \end{pmatrix}
\end{equation}
with
\begin{align}
  \tilde V^x  &:=  \frac{1}{N} \sum_{j,p,q \in \Z_N} \omega^{j(p-q)}
                 |p\>\<S^x_q| \\
              &=   \sum_{p\in \Z_N} |p\>\<S^x_p|
\,.
\end{align}
Clearly, $U^x$ can be implemented equivalently if and only if $\tilde
U^x$ can be implemented efficiently.  Therefore, if we have an
efficient quantum circuit for the transformation
\begin{equation}
  |p,x\>  \mapsto  
  \begin{cases}
    |S^x_p,x\>  & \eta^x_p > 0 \\
    |\psi^x_p\> & \eta^x_p = 0
  \end{cases}
\label{eq:qsample}
\end{equation}
where $|\psi^x_p\>$ is any state allowed by the unitarity of $\tilde{U}^x$ 
(i.e., if we can efficiently quantum sample from subset sum
solutions for legal inputs), then by running this circuit in reverse,
we can efficiently implement $\tilde U^x$, and hence the measurement.

Conversely, given the ability to implement the optimal POVM by the
measurement of $x$ followed by an efficient implementation of $U^x$,
we can solve the subset sum problem.  By running the quantum circuit
for $\tilde U^x$ in the reverse direction, we can efficiently
implement the transformation (\ref{eq:qsample}).  Suppose we are
trying to solve the subset sum problem for a legal instance $(x,t)$.
Using (\ref{eq:qsample}), we can produce the state $|S^x_t\>$, which
upon measurement gives a uniformly random subset of $x$ summing to
$t$.  On the other hand, if the instance is not legal, then we can
easily check that the output does not correspond to a subset of $x$
summing to $t$.

If we could efficiently implement the unitary operation $\tilde U^x$
for any $k=\poly(\log N)$, then we could solve the subset sum problem
efficiently even in the worst case.  Since the subset sum decision
problem is NP-complete, such an implementation seems unlikely.
However, for the purpose of solving the DHSP, it is sufficient to
consider fixed $k$ (as a function of $N$, with $\nu = k/\log N > 1$
according to Theorem~\ref{thm:threshold}) and implement the
measurement approximately.  In this case, an implementation of the
measurement only implies a solution to the average case subset sum
problem at density $\nu$, which may be considerably easier.
Conversely, to implement the measurement at density $\nu$, it is
sufficient to approximately quantum sample subset sum solutions at that
density.

The critical density $\nu \sim 1$ for the success of the optimal
measurement coincides with the critical density above which almost all
subset sum instances are legal and below which almost all subsets have
a distinct sum.  No efficient algorithms are known for the subset sum
problem at this critical density.  But for sufficiently low or high
density, the problem becomes tractable.  For densities $\nu<0.941$,
there is an efficient algorithm assuming the ability to find short
vectors in lattices \cites{Brickell:83,LO:85,Frieze:86,Coster:92}.
Unfortunately, this lattice problem seems to be difficult.  However,
using known basis reduction algorithms, this approach can be used to
efficiently solve subset sum problems with no computational
assumptions for very low density, $k < c \sqrt{\log N}$ for some
constant $c$ \cites{LO:85,Frieze:86}.

Since we require $\nu > 1$ for a solution to the DHSP, the high
density regime is more interesting for our purposes.  Until recently,
the best known result was a $\poly(k)$-time algorithm for the case
$k>cN$ for some constant $c$ \cites{CFG:89,GM:91}.  These results are
not helpful since they yield algorithms whose running times are
exponential in $\log N$.  However, Flaxman and Pryzdatek recently
showed how to produce subset sum solutions in $\poly(k)$ time with
$k=2^{O(\sqrt{\log N})}$ \cite{FP:05}.  Their result, together with
Regev's connection to the subset sum problem
(Theorem~\ref{thm:regev}), gives an alternative subexponential time
quantum algorithm for the DHSP with the same performance as
Kuperberg's algorithm.  However, it is not immediately clear whether
their algorithm can be used to quantum sample (or even to randomly
sample) from subset sum solutions, so it does not immediately provide
an implementation of the optimal measurement.

It is not inconceivable that one could find a quantum algorithm (or
even a classical one) for the subset sum problem at still lower
density, and thus find an improved algorithm for the DHSP.
Furthermore, we remark that our restriction of first measuring $x$ and
then implementing the appropriate measurement conditional on $x$,
while natural from a representation-theoretic viewpoint, is not
necessarily the best way to implement the optimal measurement.  A
direct implementation of the measurement without first measuring $x$
could in principle produce a quantum algorithm for the DHSP without
solving the subset sum problem.

\section{Discussion}
\label{sec:discussion}

In this paper we have studied the optimal measurement for
distinguishing dihedral hidden subgroup states for order two
subgroups.  Using a result of Holevo and Yuen, Kennedy, and Lax, we
proved that the pretty good measurement is optimal for this problem.
We showed that the success probability of this measurement has a
threshold around the critical density $\nu \sim 1$, and in particular,
that $\Omega(\log N)$ hidden subgroup states are necessary for the
measurement to succeed with more than an exponentially small success
probability.  We also demonstrated that the problem of determining
just the least significant bit of the answer is essentially no easier
than the full problem.  Finally, we considered the implementation of
the measurement by a quantum circuit and found that it is closely
related to the subset sum problem.  We considered the special (but
well-motivated) case in which the measurement first determines the
block $x$, and then performs the optimal POVM within that block.  For
a given number of copies of the hidden subgroup state, we showed that
this measurement can be implemented efficiently if and only if one can
quantum sample from subset sum solutions at the corresponding density.

Many open questions remain.  First, given Kuperberg's subexponential
time algorithm for the DHSP using $k=2^{O(\sqrt{\log N})}$ copies of
the hidden subgroup state, as well as the Flaxman-Pryzdatek algorithm
for finding a subset sum solution at the corresponding density, it
seems promising to look for an implementation of the optimal
measurement by quantum sampling from subset sum solutions at this
density.  As an intermediate step, it would be interesting simply to
find an algorithm for producing a subset sum solution uniformly at
random.

Of course, implementing the optimal measurement at the Kuperberg
density would not yield an improvement over previous algorithms, so it
would be more interesting to find an implementation of the optimal
measurement at still lower density.  If one pursues the natural
strategy of first measuring the block $x$, then our results show that
this approach is at least as hard as solving the subset sum problem,
in which case one could simply apply Regev's Theorem~\ref{thm:regev}.
However, as discussed above, one could consider implementing the
optimal measurement without first measuring $x$, which might give an
improved algorithm for the DHSP without yielding an algorithm for
subset sum.

Finally, it would interesting to consider optimal measurements for
other non-abelian hidden subgroup problems.  Can such measurements be
implemented efficiently in any of the cases where efficient algorithms
are already known?  Or more ambitiously, can any new quantum speedups
be found in this way?  Presumably the subset sum problem has some
analog for other groups, and such a problem might be interesting in
its own right.

\section*{acknowledgments}

We thank Carlos Mochon and Frank Verstraete for helpful discussions of
Theorem~\ref{thm:holevo}, and Abie Flaxman for correspondence regarding 
the algorithm in \cite{FP:05} and an earlier version thereof. 

AMC received support from the National Science Foundation under Grant
No.\ EIA-0086038, and was also supported in part by the Fannie and
John Hertz Foundation, by the Department of Energy under cooperative
research agreement DE-FC02-94ER40818, and by the National Security
Agency and Advanced Research and Development Activity under Army
Research Office contract DAAD19-01-1-0656.
WvD was supported in part by the Cambridge-MIT Institute and by the
Department of Energy under cooperative research agreement
DE-FC02-94ER40818.



\appendix
\section*{Appendix: Representation theory and the optimal measurement}

Many features of the hidden subgroup problem can be understood using
simple group representation-theoretic arguments.  Here we present such
arguments and demonstrate their application to the DHSP.

Two important representations of a group $\G$ for the HSP over that
group are the left and right regular representations of $\G$.  These
representations act on a Hilbert space spanned by vectors $\{|g\>\}_{g
\in \G}$ as
\begin{align}
  D_L(g_1)|g_2\>  &=  |g_1 g_2\> &  
  D_R(g_1)|g_2\>  &=  |g_2 g_1^{-1}\>
\,.
\end{align}
Viewed as representations of the group algebra, these two
representations are commutants of each other, i.e., $D_L(g_1) D_R(g_2)
= D_R(g_2) D_L(g_1)$.  The hidden subgroup states $\rho_\H$ defined in
(\ref{eq:hss}) commute with the left regular representation: $D_L(g)
\rho_\H=\rho_\H D_L(g)$ for all $g \in \G$.  Hence, via Schur's lemma
and the fact that the left and right regular representations are
commutants, it is easy to show that a general hidden subgroup state
can be expressed as
\begin{equation}
  \rho_\H  =  \frac{1}{|\G|} \sum_{h \in \H} D_R(h)
\,.
\end{equation}

The regular representations are reducible.  Let $\hat\G$ be a set of
labels for a complete set of irreducible representations of $\G$, and
for any $x \in \hat G$, let $\Gamma_x(g)$ be the $x$th irreducible
representation (irrep) matrix for the group element $g \in \G$.  Let
$d_x$ denote the dimension of the $x$th irrep.  Then there exists a
basis of the Hilbert space $\{|g\>\}_{g \in \G}$, labeled by
$|x,\ell,m\>$ with $x \in \hat\G$ and $\ell,m \in \Z_{d_x}$, such that $D_L$
and $D_R$ act as
\begin{align}
  D_L(g) &= \bigoplus_{x \in \hat\G} \Gamma_x(g) \otimes I_{d_x} &
  D_R(g) &= \bigoplus_{x \in \hat\G} I_{d_x} \otimes  \Gamma_x(g)
\end{align}
where $I_d$ is the $d$-dimensional identity matrix.  The unitary
transformation that transforms between the bases $\{|g\>\}_{g\in\G}$
and $\{|x,\ell,m\>\}_{x \in \hat\G, \ell,m \in \Z_{d_x}}$ is nothing but the
Fourier transform over $\G$,
\begin{equation}
  Q_\G
     :=  \frac{1}{\sqrt{|\G|}} \sum_{g \in \G} \sum_{x \in \hat\G}
       \sum_{\ell,m \in \Z_{d_x}} \sqrt{d_x}
       [\Gamma_x(g)]_{\ell,m}|x,\ell,m\>\<g|
\,.
\end{equation}
Here $[\Gamma_x(g)]_{\ell,m}$ is the matrix element in the $\ell$th row and
$m$th column of the $x$th irrep at the group element $g$.

If we perform the quantum Fourier transform over $\G$ on the state
$\rho_\H$, we find in the new basis
\begin{align}
  \rho_\H
     &=  \frac{1}{|\G|} \bigoplus_{x \in \hat\G}
       I_{d_x} \otimes \left( \sum_{h \in \H}\Gamma_x(h) \right) \\
     &=  \sum_{x \in \hat\G} p(x) \, \frac{I_{d_x}}{d_x} 
       \otimes \rho_{\H,x} \otimes |x\>\<x|
\,,
\end{align}
a classical mixture over the irrep label $x \in \hat\G$ with
probabilities
\begin{equation}
  p(x) := \frac{d_x}{|\G|} \sum_{h \in \H} \chi_x(h)
\end{equation}
(where $\chi_x := \tr \Gamma_x$ denotes the character of the $x$th
irrep) of a maximally mixed row state $I_{d_x}/d_x$ and the column
state
\begin{equation}
  \rho_{\H,x} 
     :=  \frac{1}{\sum_{h \in \H} \chi_x(h)} \sum_{h \in H} \Gamma_x(h)
\,.
\end{equation}
Since the row state is maximally mixed, it is clear that one learns
nothing by measuring the row index \cite{Grigni:04a}.  Thus, it is
natural to work in this basis and to discard the row state, focusing
on the column state $\rho_{\H,x}$.  This procedure corresponds exactly
to the particular form of the optimal POVM considered in
Section~\ref{sec:subsetsum}.

To see this correspondence in detail for the DHSP, consider the
irreducible representations of the dihedral group \cite{Serre:77a}.
These irreps are all either one- or two-dimensional.  The
two-dimensional irreps may be conveniently labeled by an integer $1
\le x \le \lceil{N/2}\rceil-1$, and are given by
\begin{equation}
  \Gamma_x(s^k) =
    \begin{pmatrix}
      \omega^{xk} & 0 \\
      0 & \omega^{-xk} \\
    \end{pmatrix}
     \textrm{~and~}  
  \Gamma_x(rs^k) =
    \begin{pmatrix}
      0 &\omega^{-xk}  \\
     \omega^{xk} & 0 \\
    \end{pmatrix}
\end{equation}
where $\omega:=\exp(2 \pi \i/N)$.  Notice that the irreps satisfy
$\Gamma_{-x}(g)=X \Gamma_x(g) X$ where $X$ is the Pauli matrix
\begin{equation}
  X  :=  \begin{pmatrix} 
         0 & 1 \\ 
         1 & 0 
       \end{pmatrix}
\,.
\end{equation}
When $N$ is odd, there are two one-dimensional irreps, the trivial
irrep
\begin{equation}
  \Gamma_\tau(s^k)  =  \Gamma_\tau(rs^k) =  1
\end{equation}
and the alternating irrep
\begin{equation}
  \Gamma_\sigma(s^k) = 1
    \textrm{~and~}   
  \Gamma_\sigma(rs^k) = -1
\,.
\end{equation}
When $N$ is even, there are two additional one-dimensional irreps, the
even irrep
\begin{equation}
  \Gamma_e(s^k)  =  \Gamma_e(rs^k)  = (-1)^k
\end{equation}
and the odd irrep
\begin{equation}
  \Gamma_o(s^k) = (-1)^k
    \textrm{ and }  
  \Gamma_o(rs^k) = -(-1)^k
\,.
\end{equation}

Now consider the approach to the DHSP of first performing a quantum
Fourier transform over $\G = {\mathcal D}_N$ and then measuring the
irrep index.  If the result obtained corresponds to a two-dimensional
irrep, then we can measure the row index and will randomly obtain one
of two outcomes.  We associate one of these outcomes with the irrep
label $x$ and the other with the irrep label $-x$, performing the $X$
operation on the column in the latter case.  Furthermore, we group the
trivial irrep and the alternating irrep together into a
two-dimensional space, and similarly for the even and odd irreps.
Labeling these two spaces $0$ and $N/2$, respectively, it is now easy
to see that this procedure is equivalent to the measurement procedure
outlined in Section~\ref{sec:subsetsum}, where the irrep label $x$
corresponds to the measurement of $x \in \Z_N$.

\end{document}